\documentstyle[twoside,fleqn,espcrc2]{article}

\input{psfig}

\title{D-Theory: Field Theory via Dimensional Reduction of Discrete Variables}

\author{B. B. Beard \address{Departments of Physics and Mechanical Engineering
\\ Christian Brothers University, Memphis, TN 38104},
R. C. Brower \address{Department of Physics 
\\ Boston University, Boston, MA02215},
S. Chandrasekharan$^{\ {\rm c}}$, D. Chen$^{\ {\rm c}}$, 
A. Tsapalis$^{\ {\rm c}}$ and U.-J. Wiese
\address{Center for Theoretical Physics, Laboratory for Nuclear Science and 
Department of Physics \\ 
Massachusetts Institute of Technology (MIT), Cambridge, MA 02139}}

\begin{document}

\begin{abstract}
A new non-perturbative approach to quantum field theory --- {\em D-theory} ---
is proposed, in which continuous classical fields are replaced by {\em discrete}
quantized variables which undergo {\em dimensional} reduction. The 2-d classical
$O(3)$ model emerges from the $(2+1)$-d quantum Heisenberg model formulated in 
terms of quantum spins. Dimensional reduction is demonstrated explicitly by 
simulating correlation lengths up to 350,000 lattice spacings using a loop 
cluster algorithm. In the framework of D-theory, gauge theories are formulated
in terms of quantum links --- the gauge analogs of quantum spins. Quantum links
are parallel transporter matrices whose elements are non-commuting operators.
They can be expressed as bilinears of anticommuting fermion constituents. In 
quantum link models dimensional reduction to four dimensions occurs, due to the
presence of a 5-d Coulomb phase, whose existence is confirmed by detailed 
simulations using standard lattice gauge theory. Using Shamir's variant of 
Kaplan's fermion proposal, in quantum link QCD quarks appear as edge states of a
5-d slab. This naturally protects their chiral symmetries without fine-tuning. 
The first efficient cluster algorithm for a gauge theory with a continuous 
gauge group is formulated for the $U(1)$ quantum link model. Improved estimators
for Wilson loops are constructed, and dimensional reduction to ordinary lattice
QED is verified numerically.
\end{abstract}

\maketitle

\section{Introduction}

The conventional approach to field quantization is to perform a path integral 
over classical fields. Wilson's formulation of lattice field theory maintains
close contact with perturbation theory by following this approach. This has 
obvious advantages, for example, in QCD when one wants to make contact with
continuum perturbative results. On the other hand, the close relation to
perturbative methods may imply that it is unnecessarily difficult to probe the 
non-perturbative regime. Here we propose an alternative method to quantize 
field theories. Instead of using continuous classical fields, we work with 
discrete quantized variables. In scalar field theories these variables are 
generalized quantum spins, and in gauge theories they are quantum links. It may
seem paradoxical that models with continuous symmetries can be expressed using
discrete variables. Note, however, that --- although a spin $1/2$ has only two 
discrete states --- the continuous $SO(3)$ symmetry is represented exactly. 
D-theory generalizes this structure to other symmetries.

Why should this quantization procedure be equivalent to performing a path 
integral over classical fields? A quantum spin operates in a finite Hilbert 
space. How can the full Hilbert space of a quantum field theory be recovered 
when the classical fields are replaced by analogs of quantum spins? As we will 
see, this requires specific dynamics: the dimensional reduction of discrete 
variables. This dynamics arises in quantum antiferromagnets as well as in 
quantum link models \cite{Cha97}, which are the gauge analogs of quantum spin 
systems \cite{Hor81,Orl90}. The {\em dimensional} reduction of {\em discrete} 
variables leads to a new non-perturbative {\em D-theory} formulation of quantum 
field theory \cite{Bro97a}. Quantized variables can be used e.g. to formulate 
$O(N)$, $SU(N) \otimes SU(N)$ and $CP(N)$ spin models, as well as Abelian 
$U(1)$ and non-Abelian $SO(N)$, $SU(N)$, and $U(N)$ gauge theories, in 
particular, QCD \cite{Bro97}. 

D-theory is closely related to Wilson's lattice field theory. The classical 
fields of Wilson's formulation arise as collective excitations of a large 
number of discrete D-theory variables. D-theory is defined on a much finer 
lattice than the resulting effective theory, and hence provides a microscopic 
structure underlying Wilson's theory. Our hope is that the microscopic 
structure simplifies non-perturbative calculations, and hence may be helpful
to understand the dynamics. For example, due to the discrete nature of its
degrees of freedom, D-theory can be simulated with very efficient cluster
algorithms. Also, the theory can be completely fermionized, which leads to an
interesting approach to the large $N$ limit. {\em D-theory} shares several 
interesting properties with the non-perturbative formulation of string theory, 
and can hence also be viewed as a {\em daughter} of M-theory.

This paper is based on five talks. In section 2, dimensional 
reduction of discrete variables is illustrated in the context of quantum 
antiferromagnets. Section 3 describes a finite-size scaling analysis that 
confirms dimensional reduction of the $(2+1)$-d quantum Heisenberg model to the
2-d classical $O(3)$ model using a loop cluster algorithm. In section 4,
$(4+1)$-d quantum link models are constructed and their dimensional reduction 
to ordinary 4-d Wilsonian lattice gauge theories is discussed. Section 5 
describes a numerical study of 5-d non-Abelian gauge theories, confirming the 
presence of a Coulomb phase, which is essential for dimensional reduction. In 
section 6, the D-theory formulation of full QCD is obtained by including quarks
as edge states on a 5-d slab. The first efficient cluster algorithm for a gauge 
theory with a continuous gauge group is constructed in section 7 for the $U(1)$ 
quantum link model. Finally, section 8 contains our conclusions.

\section{The 2-d $O(3)$ Model from Dimensional Reduction of Quantum Spins$\,^1$}
\footnotetext[1]{Based on a talk given by U.-J. Wiese}

Quantum antiferromagnets like $\mbox{La}_2\mbox{CuO}_4$ and
$\mbox{Sr}_2\mbox{CuO}_2\mbox{Cl}_2$ are the precursor insulators of 
high-temperature superconductors. These materials are well described by the 
2-d antiferromagnetic square-lattice spin 1/2 quantum Heisenberg model, which 
is formulated in terms of quantum spin operators $\vec S_x$ (Pauli matrices) 
with the usual commutation relations
\begin{equation}
[S_x^i,S_y^j] = i \delta_{xy} \epsilon_{ijk} S_x^k.
\end{equation}
The corresponding Hamilton operator
\begin{equation}
H = J \sum_{x,\mu} \vec S_x \cdot \vec S_{x+\hat\mu},
\end{equation}
is defined on a 2-d square-lattice. Here we consider antiferromagnets, i.e.
$J>0$. The problem has a global $SO(3)$ symmetry, because $[H,\sum_x \vec S_x]
= 0$, i.e., the Hamilton operator commutes with the total spin. The quantum 
statistical partition function takes the form
\begin{equation}
Z = \mbox{Tr} \exp(- \beta H).
\end{equation}

When $Z$ is expressed as a path integral, the inverse temperature $\beta$ 
occurs as the extent of the Euclidean time direction. In the zero temperature 
limit the quantum system resembles a 3-d classical model at infinite volume. 
The 2-d quantum Heisenberg model \cite{Bar91,Wie94,Bea96} as well as the 
experimental systems \cite{Gre94} exhibit long-range antiferromagnetic order. 
Consequently, the $SO(3)$ symmetry of the corresponding 3-d classical model is 
spontaneously broken to $SO(2)$. Then, due to Goldstone's theorem, two massless 
bosons --- in this case two antiferromagnetic magnons (or spin-waves) --- arise.
Using chiral perturbation theory, the low-energy dynamics can be described by 
effective fields in the coset $SO(3)/SO(2) = S^2$ \cite{Leu90}. Hence, the 
magnon field is a unit vector $\vec s$ --- the same field that appears in the 
classical $O(3)$ model. This is how the usual path integral variables arise 
from D-theory. Due to spontaneous symmetry breaking, the collective excitations 
of many discrete quantum spin variables form an effective continuous classical 
field.

To lowest order in chiral perturbation theory, the effective action describing
the low-energy magnon dynamics is given by
\begin{equation}
\label{spinaction}
S[\vec s] = \int_0^\beta dx_3 \int d^2x \ \frac{\rho_s}{2}
[\partial_\mu \vec s \cdot \partial_\mu \vec s + 
\frac{1}{c^2} \partial_3 \vec s \cdot \partial_3 \vec s].
\end{equation}
Here $c$ and $\rho_s$ are the spin-wave velocity and the spin stiffness. The
index $\mu$ extends over the spatial directions 1 and 2 only. At finite 
temperature $T$, the 2-d quantum system resembles a 3-d classical model with 
finite extent $\beta = 1/T$ in the Euclidean time direction. At zero temperature
spontaneous symmetry breaking implies the existence of massless Goldstone 
bosons with an infinite correlation length $\xi$. At non-zero temperature the
extent $\beta$ of the extra dimension becomes finite. Then the system appears 
dimensionally reduced to two dimensions, because $\beta \ll \xi$. However, in 
two dimensions the Mermin-Wagner-Coleman theorem prevents the existence of 
interacting massless Goldstone bosons \cite{Mer66}. As a consequence, in the 
2-d $O(3)$ model a mass gap is generated non-perturbatively. 

Is the now finite correlation length large enough for dimensional reduction 
still to occur? This question was first answered by Chakravarty, Halperin and 
Nelson \cite{Cha89}, and then studied in more detail by Hasenfratz and 
Niedermayer \cite{Has91}. They used a block spin renormalization group 
transformation that maps the 3-d chiral perturbation theory model with finite 
extent $\beta$ to a 2-d lattice $O(3)$ model. The 3-d magnon field is averaged 
over volumes of size $\beta$ in the Euclidean time direction and $\beta c$ in 
the two spatial directions. Due to the large correlation length, the field is 
essentially constant over these blocks. The averaged field is defined at the 
block centers, which form a 2-d lattice, whose spacing $\beta c$ is much 
coarser than the microscopic lattice spacing $a$ of the original quantum 
antiferromagnet. The effective action of the averaged field resembles a 2-d 
lattice $O(3)$ model formulated in Wilson's framework. However, since the 
effective coarse lattice action is obtained by exact blocking of the original 
fine lattice degrees of freedom, its lattice artifacts are entirely due to the 
microscopic lattice. This is important for the practical applicability of 
D-theory. Using the 3-loop $\beta$-function of the 2-d $O(3)$ model together 
with its exact mass-gap \cite{Has90}, Hasenfratz and Niedermayer \cite{Has91} 
generalized an earlier result of Chakravarty, Halperin and Nelson \cite{Cha89} 
to what we call the $\mbox{CH}_2\mbox{N}_2$-formula
\begin{equation}
\label{CH2N2}
\xi = \frac{e c}{16 \pi \rho_s} \exp(2 \pi \beta \rho_s)
[1 - \frac{1}{4 \pi \beta \rho_s} + {\cal O}(\frac{1}{\beta^2 \rho_s^2})].
\end{equation}
Here $e$ is the base of the natural logarithm. Since in the zero-temperature
limit $\xi$ is infinitely larger than $\beta$, the system undergoes dimensional
reduction to two dimensions when the extent of the extra dimension becomes 
large. The exponential increase of $\xi$ in the zero-temperature limit is a 
consequence of asymptotic freedom of the 2-d classical $O(3)$ model. The factor
$2 \pi$ in the exponent is the 1-loop coefficient of the corresponding 
$\beta$-function, and 
\begin{equation}
1/g = \beta \rho_s,
\end{equation}
is the inverse coupling constant of the dimensionally reduced lattice model. 
Hence, in the low-temperature limit, the 2-d antiferromagnetic quantum 
Heisenberg model provides a new non-perturbative regularization of the 2-d 
$O(3)$ model. This D-theory formulation is entirely discrete, and still has the 
continuous $SO(3)$ symmetry.

The dimensional reduction of a $(d+1)$-dimensional D-theory (in this case the 
$(2+1)$-d quantum Heisenberg model) to an effective $d$-dimensional Wilsonian 
lattice theory with lattice spacing $\beta c$ is illustrated in fig.1.
\begin{figure}[t]
\psfig{figure=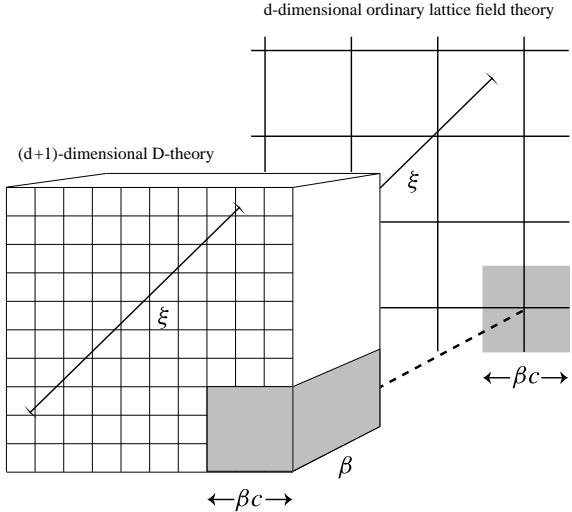,width=7.5cm}
\vspace*{-1cm}
\caption{Dimensional reduction of a D-theory: Averaging the $(d+1)$-dimensional
effective field of the D-theory over blocks of size $\beta$ in the extra 
dimension and $\beta c$ in the physical directions results in an effective
$d$-dimensional Wilsonian lattice field theory with lattice spacing $\beta c$.}
\vspace{-0.5cm}
\end{figure}
The continuum limit of the effective lattice theory is reached as 
$g = 1/\beta \rho_s \rightarrow 0$, and hence as the extent $\beta$ of the 
extra dimension becomes large. Still, in physical units of $\xi$, the extent 
$\beta \ll \xi$ becomes negligible in this limit. In the continuum limit, the 
lattice spacing $\beta c$ of the effective 2-d Wilsonian lattice $O(3)$ model 
becomes large in units of the microscopic lattice spacing of the quantum spin 
system. Hence, D-theory introduces a discrete substructure underlying Wilson's 
lattice theory. Due to exact blocking the lattice artifacts are entirely due to
the microscopic D-theory lattice.

\section{Continuous-Time Loop Cluster Simulation of the Heisenberg Model at 
Very Large Correlation Lengths$\,^2$}
\footnotetext[2]{Based on a talk given by B. B. Beard}

Soon after the discovery of high-temperature superconductivity in doped 
lamellar copper oxides it was found that the undoped compounds are 2-d spin-1/2
quantum antiferromagnets. Through experimental, theoretical, and numerical 
efforts much progress has been made in the understanding of these systems. In 
particular, detailed neutron scattering measurements \cite{Gre94} of the 
correlation length $\xi$ in the magnet $\mbox{Sr}_2\mbox{CuO}_2\mbox{Cl}_2$
were found to be in apparent good agreement with the 
$\mbox{CH}_2\mbox{N}_2$-formula eq.(\ref{CH2N2}) --- i.e. with 3-loop asymptotic
scaling of the effective 2-d lattice $O(3)$ model resulting from dimensional 
reduction. However, neutron scattering measurements on higher-spin systems 
\cite{Gre94,Nak95} reveal a striking discrepancy with the 
$\mbox{CH}_2\mbox{N}_2$-formula. The puzzle concerning the applicability of 
3-loop asymptotic scaling has recently been resolved by simulating the 
Heisenberg model at very large correlation lengths $\xi \approx 350,000 a$ 
\cite{Bea97}. 

In the context of relativistic quantum field theory, where the lattice spacing
serves as an ultraviolet cut-off that is ultimately removed, the question 
of asymptotic scaling is unphysical, because it involves the bare coupling
constant. At low temperatures a 2-d quantum antiferromagnet induces a lattice 
action for an effective 2-d $O(3)$ model with lattice spacing $a' = \beta c$, 
which is much larger than the microscopic lattice spacing $a$ of the quantum
antiferromagnet, and which is determined by the physical temperature 
$T = 1/\beta$. Hence, the question of asymptotic scaling becomes a physical 
issue. A priori, it is unclear for what values of $\xi$ one should expect 
asymptotic scaling for the effective 2-d lattice action. However, it is known 
that asymptotic scaling often sets in only at very large correlation lengths. 
For example, in the 2-d classical $O(3)$ model with the standard lattice action,
3-loop asymptotic scaling sets in at about $\xi \approx 10^5$ lattice spacings 
\cite{Car95}. Hence, one might expect that the $\mbox{CH}_2\mbox{N}_2$-formula 
works only above $\xi \approx 10^5 a' = 10^5 \beta c$. However, the induced
effective lattice action is not the standard action. Indeed, the effective 
action results from exact blocking of the chiral perturbation theory action of
the magnons. This action receives cut-off effects only from the microscopic
lattice with spacing $a$. In the limit $a \rightarrow 0$ the induced coarse 
lattice action would be a perfect action. This means that in D-theory the 
cut-off effects are due to the fine lattice spacing $a$, not due to the lattice 
spacing $a' = \beta c$ of the induced $d$-dimensional effective Wilsonian 
theory. Based on this argument, one expects that in the quantum Heisenberg 
model 3-loop asymptotic scaling sets in at $\xi \approx 10^5 a$. Indeed, this 
is observed in the numerical simulations.

How can one investigate correlation lengths in the range $\xi \approx 10^5 a$?
In the classical 2-d $O(3)$ model this was possible using finite-size scaling 
methods \cite{Kim93,Car95}. To investigate the correlation length in the quantum
Heisenberg model we use the same technique. The key observation of finite-size 
scaling is that the finite-volume correlation length $\xi(L)$ of a periodic 
system of spatial size $L \times L$ is a universal function of $\xi/L$, where 
$\xi$ is the correlation length of the infinite system. Inverting this relation 
one writes
\begin{equation}
\label{scaling}
\xi = L f(1/z) = L f(\xi(L)/L).
\end{equation}
Here $z = L/\xi(L)$ is Fisher's finite-size scaling variable, which is a
renormalization group invariant measure of the physical volume. For the 
classical 2-d $O(3)$ model the universal function $f$ has already been 
determined very precisely \cite{Kim93,Car95}. Since at low temperatures the 
quantum Heisenberg model reduces to a classical 2-d lattice $O(3)$ model, we 
can use the same universal function to deduce infinite volume results from 
finite-volume correlation length data. It is important that eq.(\ref{scaling}) 
assumes universal behavior, i.e. scaling, but not asymptotic scaling. Hence, we
can test the $\mbox{CH}_2\mbox{N}_2$-formula without bias. 

The finite-size scaling function $f$ is very sensitive to small changes in the
finite-volume correlation length $\xi(L)$. A small error in $\xi(L)$ generates 
large uncertainties in the infinite-volume correlation length $\xi$. Hence, one
needs a very accurate numerical method to determine $\xi(L)$. Fortunately, for 
the quantum Heisenberg model a very efficient loop cluster algorithm exists 
\cite{Eve93,Wie94}, which practically eliminates auto-correlations in the Monte
Carlo data. Also, it is possible to construct improved estimators which 
drastically reduce statistical errors. The only remaining systematic error is 
due to the Suzuki-Trotter discretization of Euclidean time. Recently, it has 
been realized that the discretization of Euclidean time is not necessary for 
path integrals of discrete quantum systems \cite{Bea96}. This completely 
eliminates the systematic discretization error of previous methods. Also, it 
greatly reduces storage and computer time requirements, and thus allows us to 
work at very low temperatures. 

We have performed simulations at inverse temperatures $J/T = 0.5,1.0,...,3.5$ 
measuring the correlation length directly in a large volume $L \approx 6 \xi$. 
For $J/T = 4.0,4.5,...,12$ we have used the finite-size scaling technique. The 
continuous time loop cluster algorithm was used in a single cluster version, 
and an improved estimator has been implemented for the staggered correlation 
function. Correlation lengths are extracted using the second moment method
\cite{Kim93,Car95}. In all cases at least $10^5$ measurements have been 
performed. The numerical data for the infinite-volume correlation length $\xi$ 
are compared with experimental data and the $\mbox{CH}_2\mbox{N}_2$-result in 
fig.2. 
\begin{figure}[t]
\psfig{figure=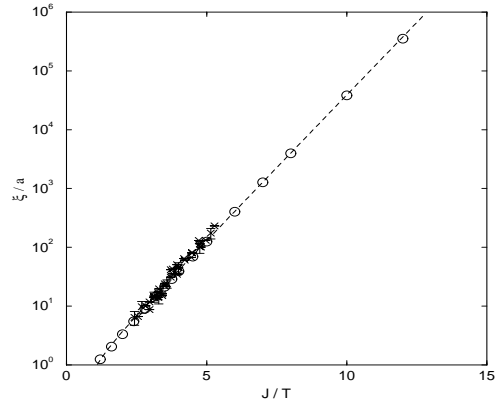,width=7.5cm}
\vspace*{-1cm}
\caption{The correlation length $\xi/a$: Experimental data (crosses) and 
numerical data (circles) are in apparent good agreement with the 
$\mbox{CH}_2\mbox{N}_2$-formula eq.(\ref{CH2N2}) (dashed line).}
\vspace{-0.5cm}
\end{figure}
The numerically accessible correlation lengths are more than three orders
of magnitude larger than the experimental ones \cite{Gre94}. 

Focusing on small effects invisible in fig.2, fig.3 shows the deviation from 
2-loop asymptotic scaling as a function of temperature. 
\begin{figure}[t]
\psfig{figure=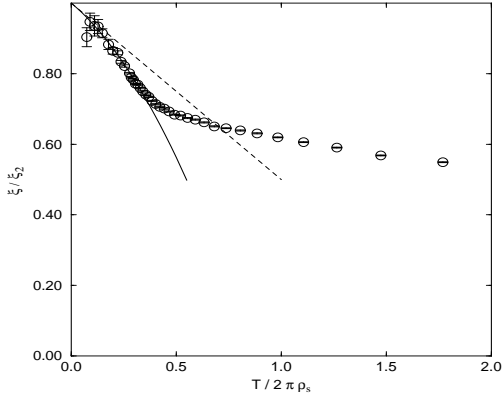,width=7.5cm}
\vspace*{-1cm}
\caption{Analysis of the correlation length $\xi$ as the asymptotic scaling
regime is approached: $\xi$ is divided by the 2-loop result
$\xi_2 = (e c/16 \pi \rho_s) \exp(2 \pi \rho_s/T)$, using the best-fit
value $\rho_s = 0.1800(5)$. The 3-loop result (dashed line) is accurate only for
$\xi \approx 10^5 a$, while the 4-loop regime (solid line) begins at
$\xi \approx 10^2 a$.}
\vspace{-0.5cm}
\end{figure}
The correlation length 
data have been fitted simultaneously with previously obtained data for the 
staggered and uniform susceptibilities in cubic \cite{Wie94} and cylindrical 
\cite{Bea96} space-time geometries. Indeed, we find that asymptotic scaling at 
the 3-loop level of the $\mbox{CH}_2\mbox{N}_2$-formula sets in at correlation 
lengths of $\xi \approx 10^5 a$, while a 4-loop fit works well already at $\xi
\approx 10^2 a$. When the massgap is included as a fit parameter, its fitted
value agrees with the exact massgap at the 2 percent level. In the light of our
numerical data, the apparent good agreement of the 
$\mbox{CH}_2\mbox{N}_2$-formula with the less accurate spin 1/2 experimental 
data at higher temperatures must be considered accidental. The large 
discrepancy between the $\mbox{CH}_2\mbox{N}_2$-formula and experimental data 
for higher-spin systems \cite{Gre94,Nak95} probably arises, because again 
3-loop asymptotic scaling sets in only at very small temperatures, which are 
inaccessible to experiments. An investigation of the spin 1 case using the 
continuous-time loop cluster algorithm is presently in progress.

The computational effort for simulating the quantum Heisenberg model is
compatible to using the Wolff cluster algorithm \cite{Wol89} directly in the 
2-d $O(3)$ model. For gauge theories, on the other hand, no efficient cluster
algorithm has been found in Wilson's formulation of the problem. Recently, the
first efficient cluster algorithm for a gauge theory with a continuous gauge
group has been constructed for the $U(1)$ quantum link model \cite{Bas97}. If
efficient cluster algorithms can be constructed also for non-Abelian quantum
link models, numerical simulations of QCD will become more accurate than the 
ones using Wilson's method.

\section{QCD as a Quantum Link Model$\,^3$}
\footnotetext[3]{Based on the talks given by R. Brower and U.-J. Wiese}

In Wilson's formulation of lattice gauge theory the parallel transporters 
$u_{x,\mu}$ are classical $SU(N)$ matrices defined on the links $(x,\mu)$ of a
4-d hypercubic lattice. The standard Wilson action takes the form
\begin{equation}
S[u] = - \sum_{x,\mu \neq \nu} \mbox{Tr} [u_{x,\mu} u_{x+\hat\mu,\nu} 
u^\dagger_{x+\hat\nu,\mu} u^\dagger_{x,\nu}].
\end{equation}
By construction, this action is invariant under gauge transformations
\begin{equation}
u'_{x,\mu} = \exp(i \vec \alpha_x \cdot \vec \lambda) u_{x,\mu}
\exp(- i \vec \alpha_{x+\hat\mu} \cdot \vec \lambda).
\end{equation}

In D-theory the Wilson action is replaced by a quantum link model Hamilton
operator
\begin{equation}
\label{Hamiltonian}
H = J \sum_{x,\mu \neq \nu} \mbox{Tr} [U_{x,\mu} U_{x+\hat\mu,\nu} 
U^\dagger_{x+\hat\nu,\mu} U^\dagger_{x,\nu}].
\end{equation}
Wilson's classical link matrices are turned into quantum link operators
$U_{x,\mu}$. They still are $N \times N$ matrices, but now their elements are 
non-commuting operators acting in a finite Hilbert space. Of course, we want to
maintain the gauge symmetry of the problem. In quantum link models, gauge 
invariance means that $H$ commutes with local generators $\vec G_x$ of gauge 
transformations at the site $x$, which obey the usual algebra
\begin{equation}
[G^a_x,G^b_y] = 2 i \delta_{xy} f_{abc} G^c_x.
\end{equation}
Under a unitary transformation $\prod_x \exp(i \vec \alpha_x \cdot \vec G_x)$,
representing a gauge transformation in Hilbert space, a quantum link variable 
transforms as
\begin{eqnarray}
U'_{x,\mu}&=&\prod_y \exp(- i \vec \alpha_y \cdot \vec G_y) U_{x,\mu} 
\prod_z \exp(i \vec \alpha_z \cdot \vec G_z) \nonumber \\
&=&\exp(i \vec \alpha_x \cdot \vec \lambda) U_{x,\mu}
\exp(- i \vec \alpha_{x+\hat\mu} \cdot \vec \lambda).
\end{eqnarray}
This relation implies
\begin{equation}
\label{GUcommutator}
[\vec G_x,U_{y,\mu}] = \delta_{x,y+\hat\mu} U_{y,\mu} \vec \lambda -
\delta_{x,y} \vec \lambda U_{y,\mu},
\end{equation}
which can be satisfied when we introduce 
\begin{equation}
\vec G_x = \sum_\mu (\vec R_{x-\hat\mu,\mu} + \vec L_{x,\mu}).
\end{equation}
Here $\vec L_{x,\mu}$ and $\vec R_{x,\mu}$ are generators of left and right
gauge transformations of the link variable $U_{x,\mu}$. They are $2(N^2-1)$
generators of an $SU(N)_L \otimes SU(N)_R$ algebra on each link. The 
commutation relations of eq.(\ref{GUcommutator}) imply
\begin{eqnarray}
\protect{[}\vec L_{x,\mu},U_{y,\nu}]&=&- \delta_{x,y} \delta_{\mu\nu} \vec\lambda
U_{x,,\mu}, \nonumber \\
\protect{[}\vec R_{x,\mu},U_{y,\nu}]&=& 
\delta_{x,y} \delta_{\mu\nu} U_{x,\mu} \vec\lambda.
\end{eqnarray}
In D-theory, the real and imaginary parts of the elements of the link matrices
are represented by $2 N^2$ Hermitean operators. Together with the operators
$\vec L_{x,\mu}$ and $\vec R_{x,\mu}$, these are $4 N^2 - 2$ generators. The
above commutation relations are those of an $SU(2 N)$ algebra, which contains
$SU(N)_L \otimes SU(N)_R$ as a sub-algebra. Of course, $SU(2 N)$ has $4 N^2 - 1$
generators, and indeed there is one more generator $T_{x,\mu}$, which obeys
\begin{equation}
[T_{x,\mu},U_{y,\nu}] = 2 \delta_{x,y} \delta_{\mu\nu} U_{x,\mu}.
\end{equation}
Consequently,
\begin{equation}
G_x = \frac{1}{2} \sum_\mu (T_{x-\hat\mu,\mu} - T_{x,\mu})
\end{equation}
generates an additional $U(1)$ gauge transformation
\begin{eqnarray}
U'_{x,\mu}&=&\prod_y \exp(- i \alpha_y G_y) U_{x,\mu} \prod_z 
\exp(i \alpha_z G_z) \nonumber \\
&=&\exp(i \alpha_x) U_{x,\mu} \exp(- i \alpha_{x+\mu}).
\end{eqnarray}
Indeed, the Hamilton operator of eq.(\ref{Hamiltonian}) is also invariant under
the extra $U(1)$ gauge transformations and thus describes a $U(N)$ lattice 
gauge theory. The symmetry can be reduced to $SU(N)$ by adding the real part of
the determinant of each link matrix to the Hamilton operator
\begin{eqnarray}
\label{Hamdet}
H&=&J \sum_{x,\mu \neq \nu} \mbox{Tr} [U_{x,\mu} U_{x+\hat\mu,\nu} 
U^\dagger_{x+\hat\nu,\mu} U^\dagger_{x,\nu}] \nonumber \\
&+&J' \sum_{x,\mu} \ [\mbox{det} U_{x,\mu} + \mbox{det} U^\dagger_{x,\mu}].
\end{eqnarray}

Due to the discrete nature of the quantum link variables, the above algebraic
structure can be expressed entirely in terms of anticommuting operators 
\begin{eqnarray}
U^{ij}_{x,\mu}&=&c_{x,+\mu}^{i \dagger} c_{x+\hat\mu,-\mu}^j, \nonumber \\ 
\vec L_{x,\mu}&=&\sum_{ij} c_{x,+\mu}^{i \dagger} \vec \lambda_{ij} 
c_{x,+\mu}^j, \nonumber \\
\vec R_{x,\mu}&=&\sum_{ij} c_{x+\hat\mu,-\mu}^{i \dagger} \vec \lambda_{ij} 
c_{x+\hat\mu,-\mu}^j, \nonumber \\
T_{x,\mu}&=&\sum_i (c_{x,+\mu}^{i \dagger} c_{x,+\mu}^i - 
c_{x+\hat\mu,-\mu}^{i \dagger} c_{x+\hat\mu,-\mu}^i).
\end{eqnarray}
Here $c_{x,\pm\mu}^{i \dagger}$ and $c_{x,\pm\mu}^i$ are creation and 
annihilation operators of colored fermions living on a link that emanates from
the site $x$ in the $\pm\mu$-direction. In quantum link models these fermions
can be viewed as constituents of the gluons, which we call rishons. The rishon 
operators obey canonical anticommutation relations
\begin{eqnarray}
\{c^i_{x,\pm\mu},c^{j \dagger}_{y,\pm\nu}\}&=&\delta_{xy} 
\delta_{\pm\mu,\pm\nu} \delta_{ij}, \nonumber \\ 
\{c^i_{x,\pm\mu},c^j_{y,\pm\nu}\}&=&
\{c^{i \dagger}_{x,\pm\mu},c^{j \dagger}_{y,\pm\nu}\} = 0. 
\end{eqnarray}
It is easy to show that the rishon number
\begin{equation}
{\cal N}_{x,\mu} = \sum_i (c^{i \dagger}_{x+\hat\mu,-\mu} c^i_{x+\hat\mu,-\mu}
+ c^{i \dagger}_{x,+\mu} c^i_{x,+\mu}),
\end{equation}
is conserved separately for each link. In rishon representation the determinant
of a quantum link operator takes the form
\begin{equation}
\mbox{det} \ U_{x,\mu} = N! c^1_{x,+\mu} c^{1 \dagger}_{x,-\mu} ... 
c^N_{x,+\mu} c^{N \dagger}_{x,-\mu}.
\end{equation}
The $U(N)$ symmetry can be reduced to $SU(N)$ via the determinant only when 
one works with exactly $N$ rishons on each link. This corresponds to working 
with the $(2N)!/(N!)^2$-dimensional representation of $SU(2N)$. Inserting the
rishon representation of $U_{x,\mu}$ in the Hamilton operator, one can describe
the quantum link dynamics as a hopping of rishons. This is illustrated in fig.4.
\begin{figure}[t]
\psfig{figure=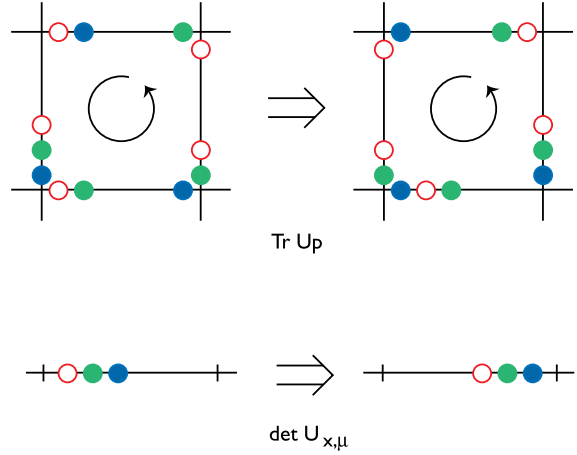,width=7.5cm}
\vspace*{-1cm}
\caption{QCD dynamics as a rishon abacus: The trace part of the Hamiltonian
induces hopping of rishons of various colors around a plaquette. The 
determinant part shifts a color-neutral combination of $N$ rishons from one end
of a link to the other.}
\vspace{-0.5cm}
\end{figure}

The partition function 
\begin{equation}
Z = \mbox{Tr} \exp(- \beta H)
\end{equation}
of a quantum link model defined on a 4-d lattice can be expressed as a 
$(4+1)$-d path integral. Note that we have not imposed the 5-d Gauss law, i.e.,
we have not included a projector on gauge invariant states propagating in the 
fifth direction. This implies that the fifth component of the non-Abelian vector
potential, $A_5 = 0$, vanishes. This is important, because it leaves us with
the correct field content after dimensional reduction. Note that the physical 
4-d Gauss law is properly imposed, because the model does contain non-trivial 
Polyakov loops in the Euclidean time direction. 

For quantum spin models, dimensional reduction arises naturally, because the
spontaneous breakdown of a continuous global symmetry provides massless 
Goldstone modes, and hence an infinite correlation length. On the other hand,
when a gauge symmetry breaks spontaneously, due to the Higgs mechanism there 
are no massless modes, and hence dimensional reduction would not occur. Also in
a confined phase the correlation length is finite. However, non-Abelian gauge 
theories in five dimensions generically have a massless Coulomb phase 
\cite{Cre79}. Based on universality, it is natural to assume the presence of a 
massless Coulomb phase also for quantum link models. Then the 5-d massless
Coulombic gluons are described by the low-energy effective action 
\begin{eqnarray}
\label{gluonaction}
S[A]&=&\int_0^\beta dx_5 \int d^4x \nonumber \\
&\times&\frac{1}{2 e^2}[\mbox{Tr} \ F_{\mu\nu} F_{\mu\nu} 
+ \frac{1}{c^2} \mbox{Tr} \ \partial_5 A_\mu \partial_5 A_\mu],
\end{eqnarray}
which is just the standard 5-d Yang-Mills action with $A_5 = 0$. The quantum 
link model is characterized by the ``velocity of light'' $c$, which is the 
analog of the spin-wave velocity in the Heisenberg model. Note that $\mu$ in
eq.(\ref{gluonaction}) runs over 4-d indices only. The dimensionful 5-d gauge 
coupling $1/e^2$ is the analog of $\rho_s$ in the spin model. At finite $\beta$ 
the above theory has a 4-d gauge invariance only, because $A_5 = 0$. At $\beta 
= \infty$ we are in the 5-d Coulomb phase with massless gluons, and hence with 
an infinite correlation length $\xi$. At finite $\beta$ the extent of the extra 
dimension becomes negligible compared to $\xi$, and the theory appears to be 
dimensionally reduced to four dimensions. However, in four dimensions the 
confinement hypothesis suggests that gluons are no longer massless. Then, as 
argued in refs.\cite{Cha97,Bro97}, a finite correlation length
\begin{equation}
\xi \propto \exp(\frac{24 \pi^2 \beta}{11 N e^2})
\end{equation}
is generated non-perturbatively. Here $24 \pi^2/11 N$ is the 1-loop 
$\beta$-function coefficient of $SU(N)$ gauge theory, and
\begin{equation}
1/g^2 = \beta/e^2
\end{equation}
is the gauge coupling of the dimensionally reduced 4-d theory. The continuum 
limit $g \rightarrow 0$ of the 4-d theory is reached when the extent $\beta$ of
the fifth direction becomes large. Like the spin model, one can consider the
dimensionally reduced 4-d theory as a Wilsonian lattice theory with lattice
spacing $\beta c$ (which has nothing to do with the microscopic lattice spacing
of the quantum link model). As before, one averages the 5-d field over cubic 
blocks of size $\beta$ in the fifth direction and of size $\beta c$ in the four 
physical space-time directions. The block centers form a 4-d space-time lattice
of spacing $\beta c$, and the effective theory of the block averaged 5-d 
Coulombic gluons resembles a 4-d Wilsonian lattice gauge theory.

Higher dimensions play an important role also in string theory. For example,
the five known superstring theories are anomaly free only in ten dimensions.
The unifying low-energy effective theory of all these models is supergravity in
eleven dimensions. The various string theories are just different 
10-dimensional reductions of the same 11-dimensional effective theory. The 
coupling constant of the 10-d string theory is related to the 11-d supergravity
coupling and to the extent of the extra dimension. Attempts to formulate string
theory beyond perturbation theory have led to M-theory, which is a microscopic 
structure underlying the non-renormalizable 11-dimensional supergravity. A 
candidate for M-theory is a matrix model \cite{Ban97}, in which the classical 
string coordinates $X_\mu$ are replaced by non-commuting operators. All this is
remarkably similar to the D-theory formulation of QCD in four dimensions. 
First, the classical gluon field $A_\mu$ is replaced by non-commuting quantum 
link operators. The low-energy effective theory of the resulting D-theory is a 
non-renormalizable 5-d non-Abelian gauge theory. Finally, just as in string 
theory, the resulting gauge coupling of QCD in four dimensions 
$1/g^2 = \beta/e^2$ is related to the 5-d gauge coupling $e$ and to the extent 
$\beta$ of the fifth direction. Due to these striking similarities, it seems 
that {\em D-theory} can be viewed as a {\em daughter} of M-theory --- the 
string theorist's candidate for the mother of all fundamental theories.

\section{Numerical Verification of Dimensional Reduction in Wilsonian $SU(3)$
Lattice Gauge Theory$\,^4$}
\footnotetext[4]{Based on a talk given by D. Chen}

The above scenario of dimensional reduction crucially depends on the presence
of a massless Coulomb phase in 5-d non-Abelian gauge theories. Early numerical
evidence for a weak coupling phase in 5-d $SU(2)$ gauge theory was presented in
ref.\cite{Cre79}. Here we extend these results to $SU(3)$ and establish that
the weak coupling phase is indeed a massless Coulomb phase \cite{Bro97b}. The 
corresponding numerical simulations are performed using Wilson's formulation of
lattice gauge theory. Via universality they should also apply to D-theory. Once 
cluster algorithms become available for non-Abelian quantum link models, the 
assumption of universality will be tested in detail.

First, 5-d pure $SU(3)$ lattice gauge theory has been simulated using the
standard Wilson action on $4^5 - 8^5$ lattices. There is a strong first order
phase transition separating the confined phase at strong coupling from a weak
coupling phase. The phase transition is at $6/e_0^2 = 4.35(15)$. The question
arises if the weak coupling phase is a massless Coulomb phase or a massive 
Higgs phase. To investigate this question, a Higgs model including an explicit 
scalar degree of freedom has been studied. In the unitary gauge, the action of 
the extended model is given by
\begin{eqnarray}
S[U]&=&\frac{6}{e_0^2} \sum_{\Box} (1 - \frac{1}{3} \mbox{Re Tr} U_\Box) 
\nonumber \\
&+&\kappa \sum_{x,\mu} (1 - \frac{1}{3} \mbox{Re Tr} U_{x,\mu}).
\end{eqnarray}
The phase diagram of the model is shown in fig.5.
\begin{figure}[t]
\psfig{figure=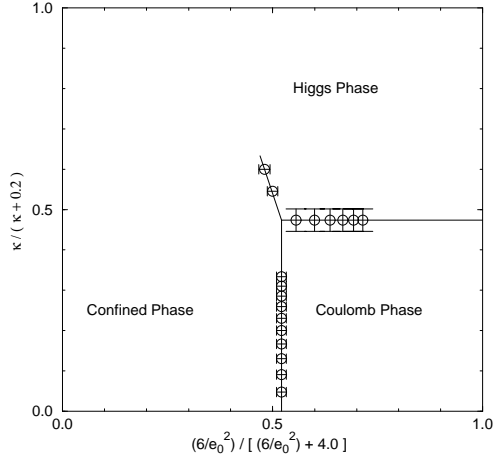,width=7.5cm}
\vspace*{-1cm}
\caption{The phase diagram of the 5-d $SU(3)$ Higgs model: The weak coupling
Coulomb phase is separated from the Higgs-confinement phase by strong first 
order phase transitions.}
\vspace{-0.5cm}
\end{figure}
The weak coupling phase of the pure gauge theory extends to small values of
$\kappa$, but is separated from the Higgs phase at large $\kappa$ by a strong 
first order phase transition. The massive Higgs and confined phases are
analytically connected.

To establish that the weak coupling phase of the pure gauge theory is indeed a
massless Coulomb phase, the static quark potential has been investigated on
a $16^5$ lattice at $6/e_0^2 = 4.6$ and 9.0 for $\kappa = 0$. Wilson loops 
$W(R,T)$ of various sizes $(R,T)$ are measured and fitted to the functional form
\begin{equation}
W(R,T) = A \exp(- e^2 F(R,T)),
\end{equation}
where $F(R,T)$ is the lattice tree-level perturbative result. Fig.6 shows the
fit for $6/e_0^2 = 4.6$. 
\begin{figure}[t]
\psfig{figure=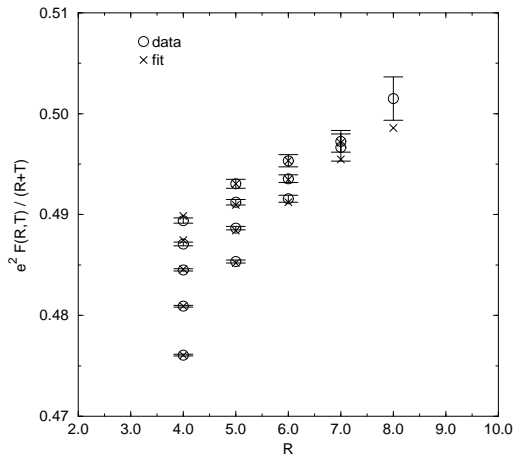,width=7.5cm}
\vspace*{-1cm}
\caption{Fit of Wilson loops: Loops of sizes $4 \times 4$ to $8 \times 8$ on a
$16^5$ lattices at $6/e_0^2 = 4.6$ are fitted to a 5-d lattice Coulomb 
potential.}
\vspace{-0.5cm}
\end{figure}
The fit parameters are $A = 1.103(5)$, $e^2 = 2.525(2)$, and 
$\chi^2/\mbox{d.o.f.} = 1.2(7)$. Similarly, for $6/e_0^2 = 9.0$ one obtains
$A = 1.0089(3)$, $e^2 = 0.8333(2)$, and $\chi^2/\mbox{d.o.f.} = 0.8(2)$. As it
should, the parameter $A$ approaches 1 in the $e_0 \rightarrow 0$ limit. Also
note that the charge renormalization $e^2/e_0^2$ drops from 1.936 at 
$6/e_0^2 = 4.6$ to 1.25 at $6/e_0^2 = 9.0$. Finally, when a massterm for the 
gauge bosons is included in the fit, its fitted value is consistent with zero. 
This confirms the presence of a massless weak coupling phase with a 5-d Coulomb 
potential $V(R) = e^2/R^2$.

To verify reduction from five to four dimensions, the 4-d finite temperature
phase transition has been investigated using 5-d pure gauge theory with 
$A_5 = 0$. The extent $\beta$ in the fifth direction generates an effective 4-d 
gauge coupling $1/g^2 = \beta/e^2$. The extent in the fourth (Euclidean time) 
direction is also finite, and corresponds to the inverse physical temperature. 
For practical purposes, it is necessary to work with asymmetric lattices with a 
coupling $6/e_0^2$ for the space-time plaquettes and a coupling $6/e_5^2$ for 
plaquettes involving the fifth direction. The simulations were done on $16^3 
\times 6 \times 2$ lattices with $6/e_0^2 = 4.6$ and $6/e_5^2 = 2.41,...,2.49$. 
The Polyakov loop in the fourth (Euclidean time) direction is shown in fig.7.
\begin{figure}[t]
\psfig{figure=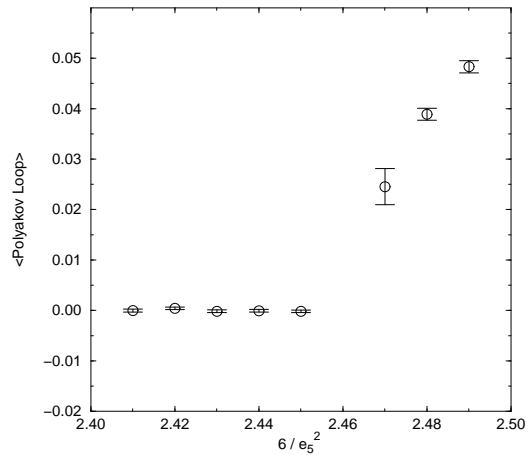,width=7.5cm}
\vspace*{-1cm}
\caption{Behavior of the 4-d Polyakov loop: Varying the coupling $6/e_5^2$ of 
plaquettes involving the fifth direction, a signal for a 
confinement-deconfinement phase transition is seen in the dimensionally reduced
4-d theory.}
\vspace{-0.5cm}
\end{figure}
One sees a clear signal for a confinement-deconfinement phase transition in
the dimensionally reduced 4-d theory. Of course, when one uses Wilson's 
method, in the pure gauge theory there is no reason to go to five dimensions.
Still, our results show that the expected dimensional reduction to ordinary 4-d
physics actually takes place, when one chooses the coupling such that the 5-d
system is in the Coulomb phase. If the quantum link model falls in the Coulomb
phase, universality implies that it also undergoes dimensional reduction. Once
cluster algorithms become available for non-Abelian quantum link models, this
scenario will be tested in detail. If powerful cluster algorithms can be 
constructed, it is likely that working with a 5-d quantum link model is more
efficient than using Wilson's method in four dimensions.

\section{Quantum Link QCD with Quarks$\,^5$}
\footnotetext[5]{Based on a talk given by R. C. Brower}

Let us now construct full QCD in the framework of D-theory. The construction 
principle of D-theory is to replace the classical action in Wilson's 
formulation by a Hamilton operator. For fermions one also replaces $\bar\psi_x$
by $\Psi^\dagger_x \gamma_5$. Thus, the quantum link QCD Hamilton operator takes
the form
\begin{eqnarray}
\label{QCDaction}
H&=&J \sum_{x,\mu \neq \nu} \mbox{Tr} [U_{x,\mu} U_{x+\hat\mu,\nu} 
U^\dagger_{x+\hat\nu,\mu} U^\dagger_{x,\nu}] \nonumber \\
&+&J' \sum_{x,\mu} \ [\mbox{det} U_{x,\mu} + \mbox{det} U^\dagger_{x,\mu}] 
\nonumber \\
&+&\frac{1}{2} \sum_{x,\mu} [\Psi^\dagger_x \gamma_5 \gamma_\mu U_{x,\mu} 
\Psi_{x+\hat\mu} \!
- \! \Psi^\dagger_{x+\hat\mu} \gamma_5 \gamma_\mu U^\dagger_{x,\mu} \Psi_x] 
\nonumber \\
&+&M \sum_x \Psi^\dagger_x \gamma_5 \Psi_x 
+ \frac{r}{2} \sum_{x,\mu} \ [2 \Psi^\dagger_x \gamma_5 \Psi_x \nonumber \\ 
&-&\Psi^\dagger_x \gamma_5 U_{x,\mu} \Psi_{x+\hat\mu}
- \Psi^\dagger_{x+\hat\mu} \gamma_5 U^\dagger_{x,\mu} \Psi_x].
\end{eqnarray}
Here $\Psi^\dagger_x$ and $\Psi_x$ are quark creation and annihilation 
operators with canonical anticommutation relations
\begin{eqnarray}
\{\Psi^{i a \alpha}_x,\Psi^{j b \beta \dagger}_y\}&=& 
\delta_{xy} \delta_{ij} \delta_{ab} \delta_{\alpha \beta}, \nonumber \\
\{\Psi^{i a \alpha}_x,\Psi^{j b \beta}_y\}&=&
\{\Psi^{i a \alpha \dagger}_x,\Psi^{j b \beta \dagger}_y\} = 0,
\end{eqnarray}
where $(i,j)$, $(a,b)$ and $(\alpha,\beta)$ are color, flavor and Dirac 
indices, respectively. The generators of $SU(N)$ gauge transformation now take 
the form
\begin{equation}
\vec G_x = \sum_\mu (\vec R_{x-\hat\mu,\mu} + \vec L_{x,\mu})
+ \Psi^\dagger_x \vec \lambda \Psi_x.
\end{equation}

The dimensional reduction of fermions is not completely straightforward. When
one uses antiperiodic boundary conditions in the extra dimension, the Matsubara
modes, $p_5 = 2 \pi (n_5 + \frac{1}{2})/\beta$, lead to a short 
${\cal O}(\beta)$ fermionic correlation length. Hence, from the point of view 
of the 4-d gluon dynamics, fermions with antiperiodic boundary conditions in 
the fifth direction stay at the cut-off and do not undergo dimensional 
reduction. When one uses periodic boundary conditions, a Matsubara mode, 
$p_5 = 0$, arises, and the quarks survive dimensional reduction, but we face 
the same fine-tuning problem that arises for Wilson fermions.

The fine-tuning  problem has been solved very elegantly in Shamir's 
variant \cite{Sha93} of Kaplan's fermion proposal \cite{Kap92}. Kaplan coupled 
5-d fermions to a 4-d domain wall, and found that a fermionic zero-mode gets 
bound to the wall. From the point of view of the dimensionally reduced theory, 
the zero-mode represents a 4-d chiral fermion. For QCD, Shamir has simplified
Kaplan's proposal by formulating the theory in a 5-d slab of finite size 
$\beta$ with open boundary conditions for the fermions at the two sides. This 
geometry fits naturally with the D-theory construction of quantum link models. 
With open boundary conditions for the quarks and with periodic boundary 
conditions for the gluons, the partition function takes the form
\begin{equation}
Z = \mbox{Tr} \langle 0|\exp(- \beta H)|0\rangle.
\end{equation}
The trace extends over the gluonic Hilbert space only. Taking the expectation
value in the Fock state $|0\rangle$, implies that there are no left-handed 
quarks at $x_5 = 0$, and no right-handed quarks at $x_5 = \beta$ \cite{Fur95}.

In the presence of quarks, the low-energy effective theory of the gluons
(with $A_5 = 0$) must be modified to
\begin{eqnarray}
\label{effaction}
S[\bar\psi,\psi,A]&=&\int_0^\beta dx_5 \int d^4x \nonumber \\
&\times&\{ \frac{1}{2 e^2} [\mbox{Tr} F_{\mu\nu} F_{\mu\nu}
+ \frac{1}{c^2} \mbox{Tr} \partial_5 A_\mu \partial_5 A_\mu] \nonumber \\
&+&\bar\psi[\gamma_\mu (A_\mu + \partial_\mu) \! + \! M \! + \! 
\frac{1}{c'} \gamma_5 \partial_5] \psi \}.
\end{eqnarray}
The ``velocity of light'' $c'$ of the quarks in the fifth direction is expected
to be different from the velocity $c$ of the gluons, because in D-theory there
is no symmetry between the four physical space-time directions and the extra 
fifth direction. This is no problem, because we are only interested in the 4-d 
physics after dimensional reduction.

Due to confinement, after dimensional reduction the gluonic correlation length 
is exponentially large, but not infinite. As explained in ref.\cite{Bro97}, the
same is true for the quarks, but for a different reason. Even free quarks pick 
up an exponentially small mass due to tunneling between the two boundaries of 
the 5-d slab. The corresponding tunneling correlation length is
$2 M \exp(M \beta)$. This suggests how D-theory can avoid the fine-tuning
problem that arises for Wilson fermions. In the chiral limit of quantum link 
QCD, the 4-d gluon dynamics takes place at a length scale
\begin{equation}
\xi \propto \exp(\frac{24 \pi^2 \beta}{(11 N - 2 N_f) e^2}),
\end{equation} 
which is determined by the 1-loop coefficient of the $\beta$-function of QCD 
with $N_f$ massless quarks and by the 5-d gauge coupling $e$. If one chooses
\begin{equation}
M > \frac{24 \pi^2}{(11 N - 2 N_f) e^2},
\end{equation}
the chiral limit is automatically approached before one reaches the continuum
limit as $\beta$ becomes large. This solves the fermion doubling problem as 
well as the resulting fine-tuning problem in the D-theory formulation of QCD.

\section{A Flux Cluster Algorithm for the $U(1)$ Quantum Link Model$\,^6$}
\footnotetext[6]{Based on a talk given by A. Tsapalis}

Cluster algorithms have been extremely successful in solving lattice field
theories with a global symmetry, like Ising and Potts models \cite{Swe87} as 
well as $O(N)$ models \cite{Wol89,Tam90} and $\Phi^4$ theory \cite{Bro89}. In 
these cases critical slowing down is practically eliminated, which allows very 
precise numerical simulations close to the continuum limit. In these cases, the 
clusters can be identified with physical excitations \cite{For72,Con80}. The 
cluster size is related to a susceptibility, which implies that the clusters
cannot grow beyond the physical correlation length. This is essential for the 
efficiency of these algorithms. For other models, like $CP(N)$ models with 
$N>1$ \cite{Jan92} and $SU(N) \otimes SU(N)$ chiral models with $N>2$, cluster 
algorithms can also be constructed. However, they turn out not to be efficient,
as discussed in Ref.\cite{Sok93}. In these cases the clusters are not related 
to a physical quantity. For lattice gauge theories the situation is even more 
difficult. Only for models with discrete gauge groups efficient cluster 
algorithms could be found \cite{Ben90,Bro90}. Despite numerous attempts, so far 
no efficient cluster algorithm has been constructed for models with continuous 
gauge groups. These attempts failed due to the presence of frustrated 
interactions and difficulties to find an efficient embedding of discrete 
variables. 

Here we construct the first efficient cluster algorithm for a model with a
continuous gauge group --- the $U(1)$ quantum link model. When formulated as a 
$(d+1)$-dimensional D-theory, this model is expected to reproduce the physics 
of the standard Wilson formulation of $d$-dimensional compact $U(1)$ gauge 
theory. For $d=4$ there is a phase transition between a strong coupling 
confined phase and a weak coupling massless Coulomb phase \cite{Gut80}. Due to 
the infinite correlation length of 5-d photons, dimensional reduction from 5 to
4 dimensions occurs when the extent $\beta$ of the additional fifth dimension 
becomes finite. As in non-Abelian quantum link models, one has 
$1/g^2 = \beta/e^2$. Dimensional reduction occurs as long as the effective 4-d
gauge coupling $g$ is weak enough to fall in the Coulomb phase of the 4-d 
theory. At stronger coupling, and hence at a smaller extent of the fifth 
direction, one enters the confined phase. Then the correlation length is 
finite, and dimensional reduction does not occur. For $d=3$ compact $U(1)$ 
gauge theory always confines \cite{Pol75,Goe82}, but the correlation length
diverges exponentially in the weak coupling limit. Then the mechanism of 
dimensional reduction is very similar to the one in QCD going from 5 to 4 
dimensions. In particular, dimensional reduction occurs only when the extent of 
the extra dimension becomes large.

An efficient cluster algorithm for the $U(1)$ quantum link model can be
constructed, because the variables are discrete even though the gauge symmetry 
is continuous. Hence, in contrast to Wilson's formulation, no embedding problem
arises. Also, the clusters are physical objects --- namely world-sheets of 
electric flux strings propagating in the additional Euclidean dimension. The 
flux cluster algorithm is a gauge analog of the loop algorithm for the quantum 
Heisenberg model \cite{Eve93,Wie94}. Like the loop algorithm, it operates 
directly in the continuum of the additional Euclidean dimension \cite{Bea96}.
Improved estimators for Wilson loops can also be constructed.

The Hamilton operator of the $U(1)$ quantum link model is defined on a 
$d$-dimensional lattice and is given by
\begin{eqnarray}
H&=&- \frac{J}{2} \sum_{x,\mu < \nu} [U_{x,\mu} U_{x + \hat \mu,\nu} 
U^\dagger_{x + \hat \nu,\mu} U^\dagger_{x,\nu} \nonumber \\
&+&U_{x,\nu} U_{x + \hat \nu,\mu} U^\dagger_{x + \hat \mu,\nu} 
U^\dagger_{x,\mu}].
\end{eqnarray}
The quantum link operators $U_{x,\mu}$ and the generators of gauge 
transformations $G_x$ can be represented as
\begin{eqnarray}
U_{x,\mu}&=&S_{x,\mu}^1 + i S_{x,\mu}^2 = S_{x,\mu}^+, \nonumber \\ 
G_x&=&\sum_\mu (S_{x - \hat \mu,\mu}^3 - S_{x,\mu}^3),
\end{eqnarray}
where $S^i_{x,\mu}$ is a spin operator associated with a link, with the
usual commutation relations
\begin{equation}
[S_{x,\mu}^i,S_{y,\nu}^j] = i \delta_{x,y} \delta_{\mu\nu} \varepsilon_{ijk} 
S_{x,\mu}^k. 
\end{equation}
The operator $S_{x,\mu}^3 = T_{x,\mu}/2$ represents the electric flux, and 
$U_{x,\mu}$ acts as a flux raising operator.

The partition function 
\begin{equation}
Z = \mbox{Tr} \exp(- \beta H)
\end{equation}
can be formulated as a $(d+1)$-dimensional path integral. Here we describe the
situation in $d=2$. Performing a checker board decomposition $H = H_e + H_o$
of the Hamiltonian into contributions of even and odd plaquettes, and
discretizing the extra dimension into $N$ intervals of size $\varepsilon$, the 
partition function takes the form
\begin{equation}
\label{Z}
Z = \lim_{\varepsilon \rightarrow 0} 
\mbox{Tr} \{[\exp(- \varepsilon H_e) \exp(- \varepsilon H_o)]^N\}.
\end{equation}
Inserting complete sets of basis states between each pair of operators yields
a $(2+1)$-dimensional lattice with $2N$ slices in the third dimension. Here we
work in an electric flux basis, and we limit ourselves to spin 1/2, such that 
the eigenvalues of $S_{x,\mu}^3$ in the slice $n$ (with $n \in \{1,2,...,2N\}$) 
at $t = \varepsilon n/2$ are limited to $e_{x,\mu,t} = \pm 1/2$.
The partition function then takes the form
\begin{equation}
Z = \prod_{x,\mu,t} \sum_{e_{x,\mu,t} = \pm 1/2} \exp(- S[e]).
\end{equation}
The action $S[e]$ is a sum of contributions from checker boarded cubes carrying
an 8-link interaction. The Boltzmann factor of a cube is determined by the 
elements of the $16 \times 16$ plaquette transfer matrix
\begin{eqnarray}
{\cal T}&=&\exp(\varepsilon \frac{J}{2} [U_{x,\mu} U_{x + \hat \mu,\nu}
U^\dagger_{x + \hat \nu,\mu} U^\dagger_{x,\nu} \nonumber \\
&+&U_{x,\nu} U_{x + \hat \nu,\mu} U^\dagger_{x + \hat \mu,\nu} 
U^\dagger_{x,\mu}]).
\end{eqnarray}
All diagonal elements of the transfer matrix are 1, except
$\langle + + - -|{\cal T}|+ + - - \rangle = 
\langle - - + +|{\cal T}|- - + + \rangle = \cosh(\varepsilon J/2) = c$, and all 
off-diagonal elements vanish, except
$\langle + + - -|{\cal T}|- - + + \rangle =
\langle - - + +|{\cal T}|+ + - - \rangle = \sinh(\varepsilon J/2) = s$. Here,
$\pm$ denotes the electric fluxes $e_{x,\mu}$, $e_{x+\hat\mu,\nu}$,
$e_{x+\hat\nu,\mu}$, $e_{x,\nu}$ around a plaquette in two adjacent time-slices.

The cluster algorithm constructs open or closed oriented surfaces, which 
represent world-sheets of electric flux strings propagating in the extra 
dimension. An update reverses the flux of all links belonging to the cluster. 
Cluster growth starts by selecting an initial link $(x,\mu,t)$ at random. The 
flux variable $e_{x,\mu,t}$ participates in two cube interactions, one at 
earlier, the other at later $t$ values. The eight links of a cube are connected
to clusters according to the following rules. For a cube configuration of 
weight $c$, with a probability $p = (c - s + 7)/8c$ each link is connected to 
its $t$-partner shifted in the extra dimension. With probability $1 - p$ the 
links are connected in two groups of four links, such that the four links with 
the same $t$ value are connected. For a configuration of weight 1, with a 
probability $q = (c - s + 7)/8$ each link is connected to its $t$-partner. With
probability $1 - q$ the links are connected in two groups of four, such that 
after flipping one group, a configuration of weight $s$ is obtained. For a 
configuration of weight $s$, with probability $r = (7c + s - 7)/8s$ the four 
links with the same $t$-value are connected. With probability $(1 - r)/7$ they 
are connected in two groups of four, such that after flipping one group, one of
the fourteen weight 1 configurations is obtained. The generated clusters form 
open or closed surfaces. The open surfaces arise, because we have not imposed 
the 5-d Gauss law for states propagating in the extra dimension. The algorithm 
is constructed such that it obeys ergodicity and detailed balance.

Due to the discrete basis of the Hilbert space, the path integral expression for
the partition function of eq.(\ref{Z}) is well-defined in the continuum of the 
extra dimension, i.e., at $\varepsilon = 0$. Taking the continuum limit of the 
cluster rules, one obtains an algorithm that operates directly in the continuum
of the extra dimension. This algorithm has been implemented along the lines of 
the continuous-time loop-cluster algorithm for the quantum Heisenberg model 
\cite{Bea96}.

To show that the clusters are directly related to physical objects, we consider
the expectation values of Wilson loops $W_{\cal C}$ --- the product of quantum 
link operators $U_{x,\mu}$ along some closed curve ${\cal C}$ on the 
$d$-dimensional lattice at a fixed value of $t$. The quantum links act as 
raising operators of electric flux. Hence, a configuration contributing to 
$\langle W_{\cal C} \rangle$ is inconsistent with the constraints in the 
configurations contributing to $Z$. Thus, it seems that just using the flux 
cluster algorithm does not provide information about Wilson loops. However, one
can imagine flipping only a part of a cluster, which is bounded by a curve 
${\cal C}$. This leads into the right sector for collecting information about 
$\langle W_{\cal C} \rangle$. In practice, one need not even perform these 
partial cluster flips explicitly, because one can extract the same information 
by examining the clusters of the original algorithm. This leads to an improved 
estimator for Wilson loops, which only receives positive contributions from 
each cluster, and shows that the clusters themselves represent physical objects.
To verify the efficiency of the algorithm we have measured autocorrelation
times, which indeed are small. A detailed study of the dynamical critical 
exponent $z$ is in progress.

The flux cluster algorithm operating in the continuum of the extra dimension
has been applied to the $(4+1)$-dimensional and $(3+1)$-dimensional $U(1)$
quantum link models. In the $(4+1)$-dimensional model the cluster size per 5-d 
volume $|C|/V$ is shown as a function of the extent $\beta$ of the fifth 
direction in fig.8. 
\begin{figure}[t]
\psfig{figure=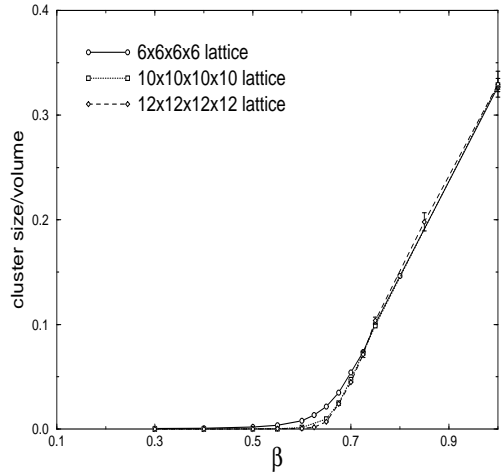,width=7.5cm}
\vspace*{-1cm}
\caption{Cluster area per 5-d volume in the $(4+1)$-d $U(1)$ quantum link model:
Above the critical extent $\beta_c \approx 0.65$ of the fifth direction, the
cluster size increases with the volume, indicating the presence of a Coulomb
phase with an infinite correlation length.}
\vspace{-0.5cm}
\end{figure}
Above $\beta_c \approx 0.65$ the cluster size is 
proportional to the volume, indicating the infinite correlation length in the
Coulomb phase. As shown in fig.9, in the $(3+1)$-dimensional model, $|C|/V$ 
goes to zero when one increases the volume, which is consistent with a confined
phase for all $\beta$.
\begin{figure}[t]
\psfig{figure=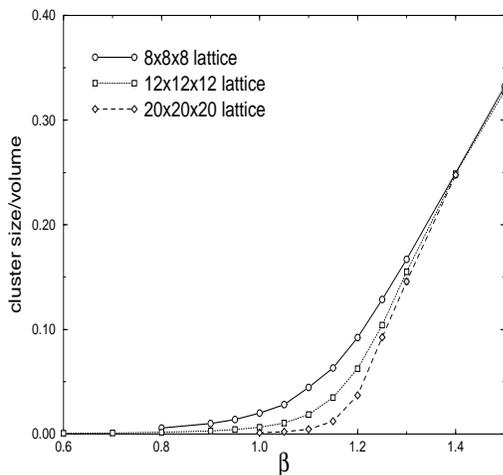,width=7.5cm}
\vspace*{-1cm}
\caption{Cluster area per 4-d volume in the $(3+1)$-d $U(1)$ quantum link model:
The data are consistent with a confined phase for all extents $\beta$ of the 
fourth direction.}
\vspace{-0.5cm}
\end{figure}
As expected, a large increase of the cluster size is observed in the large
$\beta$ limit. A detailed investigation of Wilson loops in $d=3$ and $d=4$ 
using improved estimators is in progress.

\section{Conclusions}

D-theory provides a new non-perturbative formulation of quantum field theory.
Dimensional reduction of discrete variables is a generic phenomenon that occurs
for various models, including $O(N)$, $CP(N)$, and $SU(N) \otimes SU(N)$ scalar
models, as well as Abelian and non-Abelian gauge theories, in particular QCD.
The discrete nature of the fundamental variables makes D-theory attractive, 
both from an analytic and from a computational point of view. On the analytic 
side, the discrete variables allow us to rewrite the bosonic fields in terms of 
fermionic rishon constituents. This may turn out to be useful for studying the
large $N$ limit. On the numerical side powerful cluster algorithms become 
available, which may dramatically improve numerical simulations of lattice 
field theories. A lot of work needs to be done to decide if D-theory provides a
more efficient non-perturbative quantization of field theories than Wilson's
method. 

\section*{Acknowledgments}

The work described here is supported in part by funds provided by the 
U.S. Department of Energy (D.O.E.) under cooperative research agreement 
DE-FC02-94ER40818. U.-J. W. also likes to thank the A. P. Sloan foundation for 
its support.

\end{document}